# IMPACT OF NETWORK SIZE ON THE PERFORMANCE OF INCREMENTAL LMS ADAPTIVE NETWORKS

Azam Khalili and Amir Rastegarnia
Department of Electrical Engineering,
Malayer University, Malayer, Iran
{a.khalili, a_rastegar}@ieee.org

***Abstract***—In this paper we study the impact of network size on the performance of incremental least mean square (ILMS) adaptive networks. Specifically, we consider two ILMS networks with different number of nodes and compare their performance in two different cases including (i) ideal links and (ii) noisy links. We show that when the links between nodes are ideal, increasing the network size improves the steady-state error. On the other hand, in the presence of noisy links, we see different behavior and the ILMS adaptive network with more nodes necessarily has not better steady-state performance. Simulation results are also provided to illustrate the discussions.

## I. INTRODUCTION

Distributed estimation problem arises in many applications where a set of nodes are used to estimate a parameter of interest by data collected at nodes. The environment monitoring, target localization and medical applications are just a few examples [1]. In most of these applications, the statistical information about the process of interest is not available or it varies over the time. The adaptive networks have been introduced in the literature to solve the distributed estimation problem in such cases [2-8]. The existing distributed adaptive networks can be roughly classified, based on the cooperation mode between the nodes, into incremental [2-5], diffusion [6], [7] and hierarchical [8], [9] algorithms. Our focus in this work is on incremental LMS adaptive networks where at each iteration, each node receives the prior node's local estimate, updates its local data using LMS algorithm, and finally sends it to the next node [2, 10].

The incremental adaptive networks proposed in [2-5] assume ideal links between nodes.

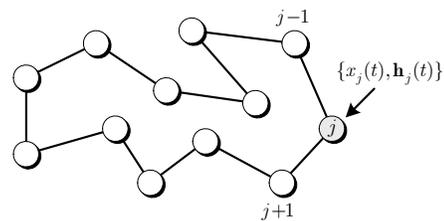

**Fig. 1 A network with incremental topology.**

However, as we have shown in [12-13], the performance of incremental adaptive network differs considerably in the presence of noisy links. In fact two key results were derived: First, noisy links leads to a larger residual MSD, as expected. Second, reducing the adaptation step size may actually increase the residual MSD.

These results are also valid for diffusion based adaptive networks with noisy link [14-18]. In this work we study the impact of network size on the performance of ILMS adaptive network. More clearly, we consider two I-LMS networks with different number of nodes ($N_1 > N_2$) and compare their performance in two different cases including (i) ideal links and (ii) noisy links. Under small step-sizes and some conditions on the data, we show that when the links between nodes are ideal, the ILMS adaptive network with $N_1$ nodes always has better steady-state performance. In the presence of noisy links, we see different behavior and the ILMS adaptive network with $N_2$ nodes necessarily has not better steady-state performance. Although the results are derived based on small step-sizes and some conditions on the data, it is true for general data distribution.

*Notation:* The symbol * is used for both complex conjugation for scalars and Hermitian transpose for matrices. We denote scalars, vector and matrices as $x$, $\mathbf{x}$ and $\mathbf{X}$ respectively. For a vector $\mathbf{x}$ and matrix $\mathbf{A}$ the weighted norm is given by $\|\mathbf{x}\|_A^2 = \mathbf{x}^* A \mathbf{x}$.

## II. BACKGROUND

### A. The Estimation Problem

Consider a network with $N$ nodes denoted by set $\mathcal{S} \triangleq \{1, 2, \cdots, N\}$ where nodes collaborate according to the incremental topology as shown in Fig. 1. At time instant $t \in \{1, 2, \ldots\}$ every node $j \in \mathcal{S}$ takes measurements from an unknown parameter of interest say $\mathbf{s}_0 \in \mathbb{R}^{M \times 1}$. The measurements taken by node $j$ and time $t$ is modeled as

$$x_j(t) = \mathbf{h}_j(t)\mathbf{s}_0 + n_j(t) \quad (1)$$

where $\mathbf{h}_j(t) \in \mathbb{C}^{1 \times M}$ denotes the *row* regression vector and $n_j(t)$ is measurement noise with variance $\sigma_{n,j}^2$ and assumed to be temporally white and spatially independent with

$$E\left[n_j^*(t_1) n_\ell(t_2)\right] = \sigma_{n,j}^2 \delta(j - \ell)\delta(t_1 - t_2) \quad (2)$$

in terms of the Kronecker delta function. The objective of the network is to estimate $\mathbf{s}_0$ in a *distributed* manner through an online learning process, where each node $j$ is allowed to collaborate only with node $j-1$. To formulate the estimation problem let denote by $\mathbf{x}$ and $\mathbf{H}$ as collected data by entire network as

$$\mathbf{x} \triangleq \begin{bmatrix} x_1 \\ x_2 \\ \vdots \\ x_N \end{bmatrix} (N \times 1), \quad \mathbf{H} \triangleq \begin{bmatrix} \mathbf{h}_1 \\ \mathbf{h}_2 \\ \vdots \\ \mathbf{h}_N \end{bmatrix} (N \times M) \quad (3)$$

Then, we can formulate the estimation task as the following unconstrained optimization

$$\begin{aligned}\mathbf{s}_0 &= \arg\min_{\mathbf{s}} E\left[\|\mathbf{x} - \mathbf{H}\mathbf{s}\|^2\right] \\ &= \sum_{j=1}^{N} E\left[|x_j(t) - \mathbf{h}_j(t)\mathbf{s}|^2\right]\end{aligned} \quad (4)$$

The optimal solution of (4) is given by the statistical information of network data $\{\mathbf{x}, \mathbf{H}\}$ as

$$\mathbf{s}_0 = \mathbf{C}_\mathbf{H}^{-1} \mathbf{C}_{\mathbf{xH}} \quad (5)$$

where

$$\mathbf{C}_{\mathbf{xH}} \triangleq E\left[\mathbf{H}^*\mathbf{x}\right], \quad \mathbf{C}_\mathbf{H} \triangleq E\left[\mathbf{H}^*\mathbf{H}\right] \quad (6)$$

In order to use (5) each node must have access to the global statistical information $\{\mathbf{C}_{\mathbf{xH}}, \mathbf{C}_\mathbf{H}\}$ which in many applications are not available or change in time. To address this issue and moreover, to enable the network to response to changes in statistical properties of data in real time, the incremental LMS adaptive network is proposed in [3].

### B. The ILMS Algorithm

The update equation in ILMS is given by

$$\begin{aligned}\hat{\mathbf{s}}_j(t) &= \hat{\mathbf{s}}_{j-1}(t) + \mu_j \mathbf{h}_j^*(t)\left(x_j(t) - \mathbf{h}_j(t)\hat{\mathbf{s}}_{j-1}(t)\right) \\ &= \hat{\mathbf{s}}_{j-1}(t) + \mu_j \mathbf{h}_j^*(t) e_j(t)\end{aligned} \quad (7)$$

where $e_j(t) \triangleq x_j(t) - \mathbf{h}_j(t)\hat{\mathbf{s}}_{j-1}(t)$. In (7) $\hat{\mathbf{s}}_j(t)$ denotes the local estimate of $\mathbf{s}_0$ at node $j$ at time $t$, $\mu$ is the step size parameter. It is shown in [12-14] that for suitably chosen step size $\mu_j$ we have

$$\lim_{t \to \infty} E[\hat{\mathbf{s}}_j(t)] = \mathbf{s}_0 + \zeta_j(\mu_j) \quad \forall j \in \mathcal{S} \quad (8)$$

In (8) $\zeta_j(\mu_j)$ denotes the bounded steady-state error term which is a function step-size parameter and

$$\lim_{\mu_j \to 0} \zeta_j(\mu_j) = 0 \quad (9)$$

When the connecting links among the node are noisy, the received local estimate at node $j$ (sent by the previous node $j-1$) becomes $\hat{\mathbf{s}}_{j-1}(t) + \mathbf{q}_j(t)$ where $\mathbf{q}_j(t)$ is channel noise term between the nodes $j-1$ and $j$ which is zero mean with covariance matrix $\mathbf{R}_{\mathbf{Q},j} = E[\mathbf{q}_j(t)\mathbf{q}_j^*(t)]$. In this case the update equation of ILMS algorithm changes to

$$\hat{\mathbf{s}}_j(t) = \hat{\mathbf{s}}_{j-1}(t) + \mu_j \mathbf{h}_j^*(t) e_j(t) + \mathbf{b}_j(t) \quad (10)$$

where in (10) $\mathbf{b}_j(t)$ represent the effects of noisy links on the update equation which is given by

$$\mathbf{b}_j(t) = \mathbf{q}_j(t) - \mu_j \mathbf{h}_j^*(t) \mathbf{h}_j(t) \mathbf{q}_j(t) \quad (11)$$

As we have shown in [10], noisy links leads to a larger residual MSE, and also, reducing the adaptation step size may actually increase the residual MSE.

## III. IMPACT OF NETWORK SIZE

In this section we study the effects of networks size (number of node) on the performance of ILMS algorithm. To this end we firstly introduce the mean-square deviation (MSD) as a metric to evaluate the performance of adaptive networks as follows

$$\text{MSD}_j = \lim_{t \to \infty} E\left[\|\tilde{\mathbf{s}}_j(t)\|^2\right] \quad (12)$$

where the weight error vector $\tilde{\mathbf{s}}_j(t)$ is defined as

$$\tilde{\mathbf{s}}_j(t) \triangleq \mathbf{s}_0 - \hat{\mathbf{s}}_j(t) \quad (13)$$

Let further denote by $\mu_1$ and $\mu_2$ the step-sizes of networks with $N_1$ and $N_2$ nodes respectively. To compare the ILMS algorithm with different number of nodes we should have

$$\mu_1 N_1 = \mu_2 N_2 \quad (14)$$

This is because the incremental algorithm uses $N$ LMS-type iterations for every measurement time. In the sequel we consider both ideal and noisy links conditions.

*A. Ideal Links*

Consider an incremental network where the measurements of every node satisfy the data model (1). The following corollaries can be deduced

1. The optimal solution $\mathbf{s}_0$ does not depend on the network size (see (5))
2. The local estimates $\mathbf{s}_j(t)$ provided by the iterative solution (7) converge to the optimal solution with a bounded error determined by $\zeta_j(\mu_j)$ (refer to (8) and (9)).
3. To have fair comparison network with larger size must have smaller step size (see (14)).

Therefore, we can conclude from the above corollaries that when the connecting links are ideal and step sizes are tuned according to (14), the network with larger network size leads to a smaller steady-state. Fig. 2 shows the impact of the network size on the performance of ILMS algorithm. We have tuned the step-size parameters as suggested by the (14). As we explained above, for equal convergence, the network with larger size has smaller steady-state error, which is given by the MSD.

It must be noted that if we do not tune the step-size parameters by (14), then network size will not affect the steady-state performance. In this case the number of nodes affects the learning of the ILMS algorithm. To show this, we derive an expression which explains how the weight error vector $\tilde{\mathbf{s}}_j(t)$ evolves in time.

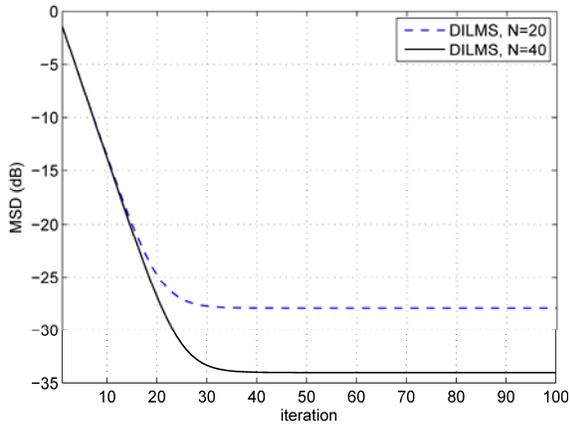

**Fig. 2 The MSD learning curves for the ILMS algorithm with different number of nodes.**

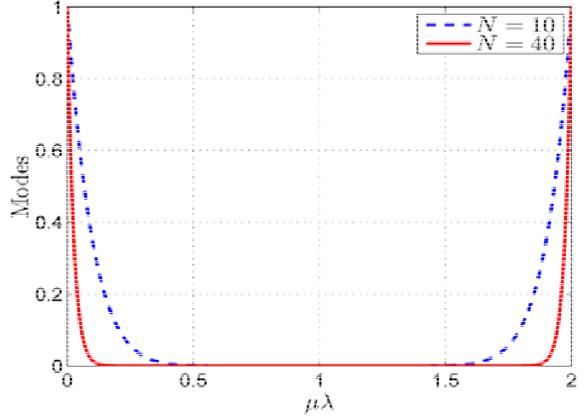

**Fig. 3 Modes of mean-convergence for ILMS algorithm with different number of nodes.**

Thus, by subtracting $\mathbf{s}_0$ from both sides of (7) we get

$$\tilde{\mathbf{s}}_j(i) = \tilde{\mathbf{s}}_{j-1}(t) - \mu_j \mathbf{h}_j^* e_j(t) \quad (15)$$

Iterating (15) and using the data model (1) we obtain that the weight-error vector evolves according to

$$E\left[\tilde{\mathbf{s}}_j(i)\right] = \left[\prod_{j=1}^{N}(\mathbf{I} - \mu_j \mathbf{C}_{\mathbf{H},j})\right] E\left[\tilde{\mathbf{s}}_{j-1}(i)\right] \quad (16)$$

where

$$\mathbf{C}_{\mathbf{H},j} \triangleq E\left[\mathbf{h}_j^* \mathbf{h}_j\right] \quad (17)$$

In order to illustrate the effect of network size on the dynamics of ILMS algorithm, we consider two network with $N = 10$ and $N = 40$ nodes. We further assume the special case of uniform regressor spatial profile, i.e., $\mathbf{C}_{\mathbf{H},j} = \mathbf{C}_{\mathbf{H}}, j \in \mathcal{S}$. The $M$ modes of convergence of the ILMS algorithm in terms of the eigenvalues $\{\lambda_m\}$ of $\mathbf{C}_{\mathbf{H}}$ is given as [10]

$$\mathcal{M} = \{(1 - \mu \lambda_m)^J\}, \quad m = 1, 2, \cdots, M \quad (18)$$

Fig. 3 shows the magnitudes of the modes of convergence for $\mathbf{C}_{\mathbf{H}} = \lambda \mathbf{I}$ and networks with $N = 10$ and $N = 40$ nodes, as a function of $\mu \lambda$. As we see the ILMS algorithm with $N = 40$ has faster convergence. Note further that the stability range for the ILMS with larger size is also wider, leading to a more robust implementation.

*B. Noisy Links*

As we mentioned in the introduction section, in the presence of noisy links reducing the adaptation step size may actually increase the residual MSD. Due to this specific property, and constraint (14), by increasing the number of nodes we see different behavior of the ILMS adaptive network. Fig. 4 shows the learning curves of the ILMS algorithm with different number of nodes when the links are noisy. We have tuned the step-sizes so that both networks have the same convergence rate. Depending on the step-size values, the network with larger number of nodes may have better, same, or even worse performance than the network with network with smaller size.

**Remark .1** It must be noted that although the observation noise profile across the network varies in general, which has an influence on the individual node performance. However, the noise profile does not affect the results of this paper. In other words, the results are valid for any distribution of data and noise.

## CONCLUSIONS

In applications where node deployment is controlled, incremental strategies are relevant since they can achieve better performance than diffusion. This was the motivation for analyzing the impact of network size on the performance of incremental networks. We considered two different cases including the ideal and noisy links. Our results revealed that when the links are ideal, increasing the network size improves the network learning and estimation performance. On the other hand, in the presence of noisy links, increasing the network size does not guarantee the performance improvement.

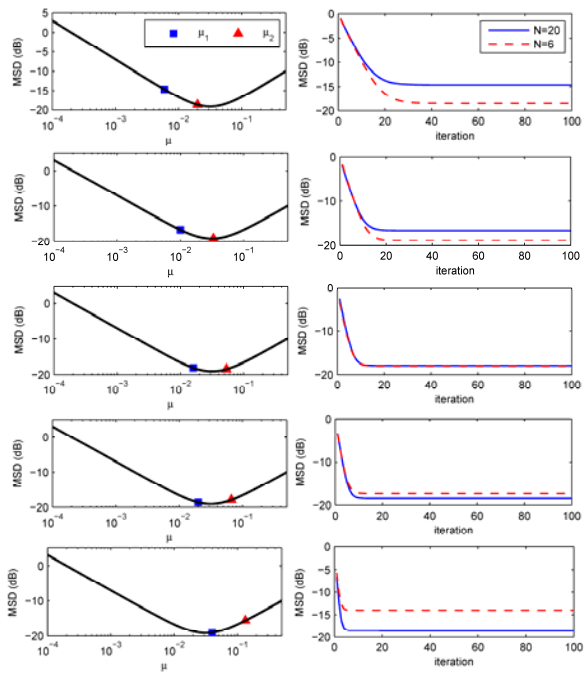

**Fig. 4** The learning behavior of the ILMS algorithm with different number of nodes. The connecting links amont the nodes are noisy.